\documentclass[12pt]{article}
\usepackage{amsfonts,latexsym}
\begin{document}
\newcommand{\de}{\delta}\newcommand{\ga}{\gamma}
\newcommand{\e}{\epsilon} \newcommand{\ot}{\otimes}
\newcommand{\be}{\begin{equation}} \newcommand{\ee}{\end{equation}}
\newcommand{\ba}{\begin{eqnarray}} \newcommand{\ea}{\end{eqnarray}}
\newcommand{\tmod}{{\cal T}}\newcommand{\amod}{{\cal A}}
\newcommand{\bemod}{{\cal B}}\newcommand{\cmod}{{\cal C}}
\newcommand{\dmod}{{\cal D}}\newcommand{\hmod}{{\cal H}}
\newcommand{\s}{\scriptstyle}\newcommand{\tr}{{\rm tr}}
\newcommand{\einsop}{{\bf 1}}
\def\oR{R^*} \def\upa{\uparrow}
\def\R{\overline{R}} \def\doa{\downarrow}
\def\oL{\overline{\Lambda}}
\def\nn{\nonumber} \def\dag{\dagger}
\def\beq{\begin{equation}}
\def\eeq{\end{equation}}
\def\bea{\begin{eqnarray}}
\def\eea{\end{eqnarray}}
\def\ve{\epsilon}
\def\si{\sigma}
\def\th{\theta}
\def\d{\delta}
\def\ga{\gamma}
\def\l{\left}
\def\r{\right}
\def\a{\alpha}
\def\b{\beta}
\def\g{\gamma}
\def\La{\Lambda}
\def\w{\overline{w}}
\def\u{\overline{u}}
\def\o{\overline}
\def\rr{\mathcal{R}}
\def\T{\mathcal{T}}
\def\N{\overline{N}}
\def\Q{\overline{Q}}
\def\L{\mathcal{L}}
\def\i{\overline{i}}
\def\j{\overline{j}}
\def\k{\overline{k}}
\def\l{\overline{l}}
\def\d{\dagger}
\newcommand{\reff}[1]{eq.~(\ref{#1})} 
%\draft

\vspace{6cm}

\begin{center}
{\Large \bf 
Reply to Schlottmann and Zvyagin}
\vskip.3in
{\large X.-Y. Ge, M.D. Gould, J. Links and H.-Q. Zhou}  
\vskip.2in
{\em Department of Mathematics, \\
The University of Queensland,
      4072, \\ Australia
     } 
      \end{center}

      \vskip 2cm
%      \begin{abstract}
%      \end{abstract}          

\vskip.3in
\begin{abstract}
This is a reply to a comment by P. Schlottmann and A.A. Zvyagin.
\end{abstract}

%\vfil\eject

%************************** Text Begins here ******************************

\def\a{\alpha}
\def\b{\beta}
\def\d{\dagger}
\def\e{\epsilon}
\def\g{\gamma}
\def\k{\kappa}
\def\l{\lambda}
\def\o{\omega}
\def\t{\tilde{\tau}}
\def\s{S}
\def\D{\Delta}
\def\T{{\cal T}}
\def\TT{{\tilde{\cal T}}}
% Shorthands for \begin{equation} and the like

\def\beq{\begin{equation}}
\def\eeq{\end{equation}}
\def\bea{\begin{eqnarray}}
\def\eea{\end{eqnarray}}
\def\ba{\begin{array}}
\def\ea{\end{array}}
\def\no{\nonumber}
\def\le{\langle}
\def\re{\rangle}
\def\lt{\left}
\def\rt{\right}
\def\oR{R^*} \def\upa{\uparrow}
\def\R{\overline{R}} \def\doa{\downarrow}
\def\oL{\overline{\Lambda}}
\def\nn{\nonumber} \def\dag{\dagger}
\def\e{\epsilon}
\def\si{\sigma}
\def\th{\theta}
\def\de{\delta}
\def\ga{\gamma}
\def\l{\left}
\def\r{\right}
\def\a{\alpha}
\def\b{\beta}
\def\g{\gamma}
\def\La{\Lambda}
\def\w{\overline{w}}
\def\u{\overline{u}}
\def\o{\overline}
\def\rr{\mathcal{R}}
\def\T{\mathcal{T}}
\def\N{\overline{N}}
\def\Q{\overline{Q}}
\def\L{\mathcal{L}}
\def\i{\overline{i}}
\def\j{\overline{j}}
\def\k{\overline{k}}
\def\l{\overline{l}}
\def\d{\dagger}

In their comment \cite{sz}, Schlottmann and Zvyagin raise several issues
regarding the nature of integrable impurities in one-dimensional quantum
lattice models, and claim to expose false statements in our recent work 
\cite{gglz}. In order to address these issues in a pedagogical manner,
we feel that it is appropriate to discuss these questions in terms of
models based on the $gl(2|1)$ invariant solution of the Yang-Baxter 
equation. However it is important from the outset to make it 
clear that the arguments we
will present below are general and apply to other classes of models. 

The first point that we would like to make is that it is claimed in \cite{sz}
that there are two approaches to the algebraic Bethe ansatz. However, 
it appears to us that approach (i) described in item (2) of \cite{sz} is the 
{\it co-ordinate} Bethe ansatz and we are completely bemused why this 
should be referred to as an algebraic Bethe ansatz. 
In the co-ordinate Bethe ansatz approach one starts with a prescribed
Hamiltonian and then solves the Schr\"odinger equation to get the 
two-particle scattering matrices and the particle-impurity scattering
matrix. Together these scattering matrices form the monodromy matrix. 
It is impossible to infer the Hamiltonian from such a monodromy matrix.
In this context we do not feel that item (1) of \cite{sz} answers our query
about the existence of the impurity monodromy matrix in the 
algebraic approach (ii) of item (2) in \cite{sz}. 
Hereafter we will focus our attention to this case.  
  
The solution of the Yang-Baxter equation 
\beq R_{12}(u-v)R_{13}(u)R_{23}(v)=R_{23}(v)R_{13}(u)R_{12}(u-v)
\label{ybe} \eeq  
associated with the Lie superalgebra $gl(2|1)$ is an operator 
$R(u)\in {\rm End}\, (V\otimes V)$, where $V$ is a three-dimensional
${\mathbb Z}_2$-graded space with one bosonic and two fermionic degrees
of freedom. Explicitly, this operator takes the form   
\beq R(u)=u.I\otimes I+P\label{rmat} \eeq  
where $P$ is the ${\mathbb Z}_2$-graded permutation operator. 
For the purposes of constructing integrable one-dimensional quantum
systems on a closed lattice it is usual to introduce the the Yang-Baxter
algebra with elements ${\hat A}_{ij}(u)$, $i,j=1,2,3$. Putting these
operators as the elements of a $3\times 3$ matrix $A(u)$, the algebraic 
relations amongst these operators are determined by the requirement 
\beq R_{12}(u-v)A_1(u)A_2(v)=A_2(v)A_1(u)R_{12}(u-v). \label{yba} \eeq  
It is apparent from eq. (\ref{ybe}) that the $R$-matrix (\ref{rmat}) 
provides a representation of the Yang-Baxter algebra. The tensor product
representation 
\beq T(u)=R_{0N}(u)......R_{02}(u)R_{01}(u), \label{mono} \eeq 
called the monodromy matrix, is also a representation of the Yang-Baxter
algebra. 
The transfer matrix $t(u)$ is defined as the supertrace of the monodromy
matrix, viz.  $t(u)={\rm str}_0 T(u)$ and the Hamiltonian is the logarithmic
derivative of the transfer matrix evaluated at $u=0$, viz.  
$H=t^{-1}(0)t^\prime (0) $, where the prime indicates the derivative
with respect to the spectral parameter $u$.  
For the present case this yields the supersymmetric $t-J$ model as
demonstrated in \cite{ek,fk}. 

In order to incorporate an integrable impurity into the model, 
one looks for a different representation of $A(u)$, say $L(u)$, and then   
constructs the impurity monodromy matrix 
\beq T(u)=R_{0N}(u)......R_{02}(u)L_{01}(u-\theta) \label{impmono} \eeq 
with the transfer matrix and Hamiltonian defined as above. We agree with
\cite{sz} that 
\begin{itemize} 
\item the position of the impurity Lax operator $L(u-\theta)$ in
the chain is inconsequential (here we put it in the first site for
convenience) 
\item that the parameter $\theta$, which plays
the role of a coupling parameter, can be chosen arbitrarily 
\item 
multiple impurity Lax operators can be placed in the monodromy matrix,  
each with an independent coupling $\theta_i$, and that in this case the
position of the impurity Lax operators has no effect on the Bethe ansatz
equations. 
\end{itemize} 

We also agree with \cite{sz} that when one extends this approach to the
case of an open chain, which is achieved by the additional requirement
that there is a solution to the reflection equations, this situation
persists. Any impurity Lax operator $L(u-\theta)$ for the closed chain is
also an impurity Lax operator for the open chain, its position in the chain
is immaterial regarding the Bethe ansatz solution, the coupling $\theta$
is arbitrary and the solution for the 
extension to multiple impurities with independent
couplings is independent of the position of  the impurities in the
chain. The essential feature which seems to be completely missed by the
authors of \cite{sz} is that the converse is not necessarily true! There
exists classes of boundary impurities, which are obtained via solutions
of the reflection equations, which have no analogue in the corresponding
closed chain case. This is one of the primary results that we have tried to
convey in \cite{gglz} for the $q$-deformed $t-J$ model. The case of the
supersymmetric $t-J$ model (with $q=1$) 
was studied in \cite{zglg}, which is relevant
to our discussions here. Our results show the existence of boundary Kondo 
impurities for the supersymmetric $t-J$ model which have no analogue in
the closed chain case. We will return to this point later. 

Next we will look, in closer detail, at the types of impurity 
Lax operators
$L(u)$ which exist for the present case of the supersymmetric 
$t-J$ model. We do not claim that the
present classification is complete. 
One class takes the generic form 
\beq L(u)=u.I\otimes I +\sum_{i,j}^3 (-1)^{[i]} e^j_i\otimes 
\pi(E^i_j) \label{lop} \eeq 
where $e^j_i$ is the matrix acting on $V$ with 1 in the $(i,\,j)$
position and zeroes elsewhere and 
$\pi$ is an arbitrary representation of the $gl(2|1)$ generators 
$E^i_j$ satisfying the commutation relations 
$$[E^i_j,\,E^k_l]=\delta^k_jE^i_l-(-1)^{([i]+[j])([k]+[l])}\delta^i_l
E^k_j. $$ 
Above, $[i]$ is the grading index so that $[i]=0$ if $i$ is a bosonic
label and $[i]=1$ if $i$ is a fermionic label. Below we will adopt the
convention $[1]=[2]=1,\,[3]=0$. Choosing the fundamental
representation in eq. (\ref{lop}) for the $gl(2|1)$ generators with 
$\theta=0$ yields the
$R$-matrix (\ref{rmat}). The case of non-zero $\theta$ gives rise to 
an impurity Lax operator which was studied in 
\cite{b}. Other impurity Lax operators can be obtained by choosing
different representations for $gl(2|1)$. 
For the case of the dual to the fundamental 
representation this yields the impurity model studied in \cite{lf,ar}.
Choosing the four-dimensional representations yields the impurity model
of \cite{bef,flp}. According to \cite{sz}, these models belong to the
class (ii) described in item (2), in which there are two levels of Bethe
ansatz equations corresponding to charge and spin degrees of freedom. 
However, it is claimed in item (4) of \cite{sz} that ``the effect of the
impurity matrix in the charge sector (first level Bethe ansatz) {\it
changes} the commutation relations in the spin sector (second
level).'' This statement is completely at odds with the explicit studies
of the Bethe ansatz solutions conducted in \cite{lf,ar,bef,flp,marcio} 
and is
also not supported by the findings of \cite{frank} where the algebraic
Bethe ansatz was performed for the $gl(2|1)$ invariant $R$-matrix 
in the most general context, with the only assumption being made the
existence of a reference state on which the action of the 
monodromy matrix takes an upper triangular form. 
This raises a serious doubt over the validity of the Bethe ansatz
solution presented in Appendix A of \cite{z}, which we have already discussed
in the introduction of \cite{gglz}. 
Some concerns about the Bethe ansatz treatments of one of the authors
of \cite{sz} have been raised by other researchers 
\cite{ss,kundu,kc,egr,gs,a} 
and we believe that
a similar lack of appreciation of the subtleties involved has caused the
current misunderstanding regarding our work. 
Even worse, in (5) of \cite{sz} the
authors attempt to counter our criticisms by claiming the operator
${\hat A}_{21}(u)$ to be a ``raising operator''.  
Appendix A of \cite{z} claims that the operators
${\hat A}_{12}(u)$, ${\hat A}_{13}(u)$, ${\hat A}_{21}(u)$ and 
${\hat A}_{23}(u)$
all vanish on the reference state. If ${\hat A}_{21}(u)$ is a raising
operator then ${\hat A}_{12}(u)$, ${\hat A}_{13}(u)$ and
${\hat A}_{23}(u)$ are lowering operators. The fact that they each
vanish on the reference state indicates that the reference state is a
``lowest weight state'', rather than a ``highest weight state''.  
Clearly, a proper treatment does not assume that the raising operator ${\hat
A}_{21}(u)$ 
vanishes on a lowest weight state. These assumptions were
not made in the Bethe ansatz treatments of \cite{lf,ar,flp,marcio} 
(\cite{bef} is an exception where such an assumption is valid) and indeed 
\cite{frank} goes into a detailed discussion about this very issue in
the general context.    

We believe also that this assumption of \cite{z} is inconsistent with the
transfer matrix eigenvalue (A1) presented therein. 
Here we give our argument
in detail.
\def\vac{\left|\Psi\right>}
Let $\vac$ denote the reference state for the algebraic Bethe ansatz
procedure. It is assumed in \cite{z}
that the reference state is an eigenstate of
the diagonal operators ${\hat A}_{ii}(u)$, viz.
\beq {\hat A}_{ii}(u)\vac=\lambda_i(u)\vac, \label{ev} \eeq
and
\beq {\hat A}_{21}(u)\vac={\hat A}_{12}(u)\vac=0. \label{van} \eeq
The explicit commutation relations between the operators
${\hat A}_{ij}(u),\,i,j=1,2$
can be conveniently expressed as \cite{kr}
$$(u-v)[{\hat A}_{ij}(u),\,{\hat A}_{kl}(v)]
={\hat A}_{il}(u){\hat A}_{kj}(v)-{\hat A}_{il}(v){\hat A}_{kj}(u)
$$
and in particular we have
\beq (u-v)[{\hat A}_{12}(u),\,{\hat A}_{21}(v)]
={\hat A}_{11}(u){\hat A}_{22}(v)-{\hat A}_{11}(v){\hat A}_{22}(u)
. \label{comm} \eeq
Acting (\ref{comm}) on $\vac$ immediately shows, through use of
(\ref{ev},\ref{van}), that
$$\lambda_1(u)\lambda_2(v)= \lambda_1(v)\lambda_2(u)$$
for all values of the parameters $u,v$, which immediately implies that
$$\lambda_2(u)=c\lambda_1(u) $$
for some constant $c$ independent of $u$. The constant will be unity if
there is spin reflection symmetry.
The action of the transfer matrix $t(u)$ on $\vac$ thus reads
$$t(u)\vac = \left(\lambda_3(u)-(1+c)\lambda_1(u)\right)\vac $$
which is inconsistent with the result
obtained by setting $M=N=0$ in (A1) of \cite{z}.

To further confuse the issue, item (5) of \cite{sz} mistakenly identifies 
the parameter $l$ introduced in the introduction of \cite{gglz}
as characterizing the symmetry of the impurity Lax operator. It is in fact 
a label for a class of atypical representations of the $gl(2|1)$ superalgebra.
Applying such a representation to eq. (\ref{lop}) produces an impurity 
Lax operator which bears resemblence to that 
alluded to in Appendix B of \cite{z}. It should be clear in this instance 
why the operator ${\hat A}_{21}(u)$ cannot vanish on the 
(lowest weight) reference state, 
since ${\hat A}_{21}(u)$ is a spin raising operator, and the ``higher spin'' 
atypical representations
of $gl(2|1)$, which are being used for the local impurity Hilbert space,
do not admit a spin singlet.   
A description and discussion of the applicability of these
representations is given in \cite{f}. 

Item (6) of \cite{sz} provides a misrepresentation of our claims. 
We agree that it is ``incorrect that impurities can only modify the first
level Bethe ansatz equations.'' The point that we make in the introduction 
of \cite{gglz} is that {\it if} ${\hat A}_{12}(u)$ {\it and} 
${\hat A}_{21}(u)$ 
{\it both} vanish on the reference state (as is assumed in \cite{z}), {\it then}
the second level Bethe ansatz equations will be unchanged (contradicting
the final results presented in \cite{z}). 

We now return to the issue of impurity Lax operators. Another class
can be obtained by taking the Lax operator
associated with a subalgebra of the symmetry of the $R$-matrix. For
example, in the present $gl(2|1)$ invariant case there is a natural
$gl(2)$ subalgebra which has the Lax operator 
\beq L(u)=u.(e_1^1+e^2_2)\otimes I 
-\sum_{i,j}^2  e^j_i\otimes
\pi(F^i_j) \label{gl2lop} \eeq
where the operators $F^i_j$ satisfy the $gl(2)$ commutation
relations. 
This impurity operator can be used in (\ref{impmono}) to construct an
impurity model which would appear to describe a Kondo type impurity,
except for one important facet. The operator (\ref{gl2lop})  
is singular, which in turn makes the transfer matrix singular, and does
not yield a well defined Hamiltonian in terms of the logarithmic
derivative of the transfer matrix. 
It is for this reason
that we looked to the reflection equations for the study of Kondo 
impurities in the $t-J$ model in \cite{zglg}. 

Finally we turn to the pertinent results of our works
\cite{gglz,zglg}. In these papers we solve the reflection equations for
the purpose of introducing Kondo impurities in the supersymmetric
$t-J$ model and the $q$-deformed version, in a fashion 
which permits a well 
defined Hamiltonian. By this approach, the impurity interaction is
purely of the Kondo type in the sense that there are no charge degrees
of freedom, in spite of the protestations of \cite{z} that this
situation is not possible. The argument of \cite{z} that charge degrees
of freedom should be present for the impurity site, 
on the basis of symmetry, is wrong because 
it is possible to break symmetry at the boundary and maintain
integrability, just as in the case of boundary scalar fields (cf.
\cite{dn} for a discussion of this point in relation to integrable models with 
broken $U_q(gl(n))$ symmetry due to boundary interactions). In
\cite{zglg} the symmetry of the Hamiltonian is broken from 
$gl(2|1)$ to $gl(2)$ due to the inclusion of the boundary interactions. 

Solutions of the reflection
equation are specific to the construction of open chain models and there
is no justification to expect that such integrable boundary impurities 
have an analogue in the bulk case. 
An attempt to do so for the model of \cite{zglg} yields a singular 
open chain transfer matrix analogous to that discussed above for the
periodic case, 
for which the Hamiltonian cannot be 
defined. We have argued vehemently in the
conclusion of \cite{zglg} that the boundary impurity $K$-matrices 
are operator valued and 
do not arise as the ``dressing'' of a scalar $K$-matrix. This indicates
that these impurities are integrable only when situated on the boundary
and do not have a bulk counterpart. (The situation really is no different 
to the case of integrable boundary scalar fields which also cannot 
be moved into the bulk. It is also true of model I in \cite{ws} studying
a spin ladder model with defect in the rung coupling, 
where integrability holds only
when the defect is on the boundary.) It is important to emphasize that 
our construction is entirely different from that discussed in \cite{zj} 
where the $K$-matrix is ``the ordinary $c$-number reflection matrix 
of a free boundary sandwiched between two forward scattering impurity 
matrices.'' This conclusion was confirmed by 
\cite{fs} who developed the ``projection method'' for the construction
of such impurities in a general context. Application to the specific
case of the $t-J$ model was studied in \cite{bf}.  Furthermore, the case of
boundary impurities in the $q$-deformed $t-J$ model was independently
studied in \cite{wadati,fw} and also confirms the validity of our approach.

%\newpage
%\vskip.3in

\end{document}